\documentclass[aps,prc,preprintnumbers,showpacs,superscriptaddress,twocolumn,floatfix]{revtex4-1}

\usepackage{graphicx}
\usepackage{dcolumn}
\usepackage{bm}
\usepackage{epsf}
\usepackage{epsfig}
\usepackage{amssymb}
\usepackage{verbatim}
\usepackage{amsmath}
\usepackage{longtable}
\usepackage[utf8]{inputenc}
\usepackage{helvet}
\usepackage[polish,english]{}
\usepackage{subdepth}
\usepackage{tikz}
\usepackage{rotating}
\usepackage{footnote}
\usepackage{threeparttable}
\usepackage{color}
\usepackage{chngpage}
\usepackage{array}
\usepackage{booktabs}
\usepackage{tabularx}
\usepackage{lipsum}
\begin{document}

\title{Large-scale shell-model calculations on the spectroscopy of  $N<126$ Pb isotopes}%

\author{Chong~Qi}
\affiliation{Department of Physics, KTH Royal Institute of Technology, 10691 Stockholm, Sweden}
\email{chongq@kth.se}
\author{L.Y. Jia}
\affiliation{
Department of Physics, University of Shanghai for Science and Technology, Shanghai 200093, China}
\author{G.J. Fu}
\affiliation{School of Physics Science and Engineering, Tongji University, Shanghai 200092, China}
\date{\today}%

\begin{abstract}
Large-scale shell-model calculations are carried out in the model space including neutron-hole orbitals $2p_{1/2}$, $1f_{5/2}$,
$2p_{3/2}$, $0i_{13/2}$, $1f_{7/2}$ and $0h_{9/2}$ to study the structure and electromagnetic properties of neutron deficient Pb isotopes. An optimized effective interaction is used. Good agreement between full shell-model calculations and experimental data is obtained for the spherical states in isotopes $^{194-206}$Pb. The lighter isotopes are  calculated with an importance-truncation approach constructed based on the monopole Hamiltonian. The full shell-model results also agree well with our generalized seniority and nucleon-pair-approximation truncation calculations.
The deviations between theory and experiment concerning the excitation energies and electromagnetic properties of low-lying $0^+$ and $2^+$ excited states and isomeric states may provide a constraint on our understanding of nuclear deformation and intruder configuration in this region.
\end{abstract}
\maketitle

\section{Introduction}
The structure of neutron-deficient lead isotopes with $N<126$ has been one of the most active subjects of nuclear physics. One prominent example is  the study of the systematics of the low-lying $0^+$ states and the possible coexistence of states with different shapes in nuclei around the neutron mid-shell nucleus $^{186}$Pb  for which substantial experimental and theoretical efforts have been devoted from different perspectives (for reviews see Refs. \cite{And00,Hey11,Nom15,Hey83,Woo92,Ab90,Julin01} and references therein) both experimentally \cite{And00,Pak05,Pak07,Dup00,Bree14,Rah10,Witt07,Bal11,Pak09,Gra08,Wil10,Wrz16} and theoretically \cite{Ben89,Smi03,Jiao15,Nom12,Hey11,Yao13,Egi,Rod04,Fra,Hell08,Dug03,Bender04,Del14}. Another interesting aspect is the gradual increase in empirical pairing gaps  in nuclei in that region when leaving the $N=126$ shell closure. It may indicate a reduction of the two-neutron correlation which can have a fundamental influence on nuclear $\alpha$ decays \cite{And13,Qi14} and two neutron transfer reactions \cite{Tak83}.
Moreover, 
there has been a long history studying the structure and electromagnetic properties of the spherical shell-model states in these Pb isotopes \cite{Blo65}.
In particular, significant efforts have been made measuring the static quadrupole moments and magnetic moments (g factors) of the isomeric $12^+$, $13/2^+$ and $33/2^+$ states in Pb isotopes \cite{zyw81,Nak72,Young75,Sten83,Sten85,Ruv86,Lin76,Blo93,Ion04,Ion07,Kmi10,Rou77}, which are supposed to be spherical and are related to the coupling of neutrons in the orbital $0i_{13/2}$.
Recent measurement on the quadrupole moments of the co-existing $11^-$ states are reported in Refs. \cite{Ion07,Vyv02,Vyv04}.
Systematics of the spherical states and the proposed co-existing deformed states in Pb isotopes can be found in Fig. 3 in Ref.
\cite{Rah10} and Fig. 7 in Ref. \cite{Hell08}. Those deformed states can be described well by 
 collective models like the interacting boson model \cite{Hell08,Nom12} which, however, has limited power in describing the spherical states.
 
The full configuration interaction
shell model has been successful in explaining many properties of the nuclear many-body system and may be expected to provide additional information on the structure of above nuclei from a microscopic perspective. Similar configuration interaction approaches also play an important role in the description of other quantum many-body systems including quantum chemistry and atomic and molecular
physics. 
However, the application of the shell model is highly restricted since the size of the configuration space increases dramatically
with the number of particles and orbitals. The study of those mid-shell nuclei was far beyond the reach of shell-model calculations.
Actually, only nuclei around $^{208}$Pb with a few valence particles (holes) were considered in most existing shell-model calculations for Pb isotopes \cite{Cor98,Lei10,Xu09,Sha12} and approximation methods have to be employed in other cases.
In Refs. \cite{Har76,San93,Cen97}, the low-lying structure of the odd-$A$ Pb isotopes were discussed in terms of one- and three-quasiparticle states. 
The two and four quasiparticle excitations in even Pb isotopes were discussed in Ref. \cite{Pom90}.

In this paper we present state-of-the-art shell-model calculations for Pb isotopes with $N<126$ by taking advantage of the significant progress that have been made in developing efficient diagonalization algorithm. Full model space calculations are done for the isotopes $^{194-206}$Pb. An importance-truncation technique is developed to study the structure of the lighter Pb isotopes. We have also carried out generalized seniority model and pair-truncated shell model calculations for above nuclei. The results are compared with those predicted by the shell model.
We hope that the long Pb isotopic chain can provide a critical test to our understanding of the shell-model effective interaction. The full shell model calculation will also be an important benchmark for other approximation methods and pair-truncated and seniority-truncated shell-model calculations.

\section{Model and monopole-based truncation}
The full configuration interaction shell model aims to construct the wave function as
a linear expansion of all possible anti-symmetric Slater
determinants within a given model space.  The model space is usually defined by including single-particle orbitals near the Fermi surface.
In general, the first step for a shell model calculation is to classify the bases in terms of ``partition'' which means a set of configurations with same definite number of particles in each orbit. Then the basis wave functions in each partition can be constructed within the so-called \emph{j-j} coupled
scheme \cite{oak} or the uncoupled M-scheme \cite{Whitehead}. In the latter case only the additive $M$ is a good quantum number.  The M-scheme is the \textit{de facto} standard approach for large-scale shell model calculations due to its simplicity.

\begin{figure}  
\begin{center}
\includegraphics[width=0.39\textwidth]{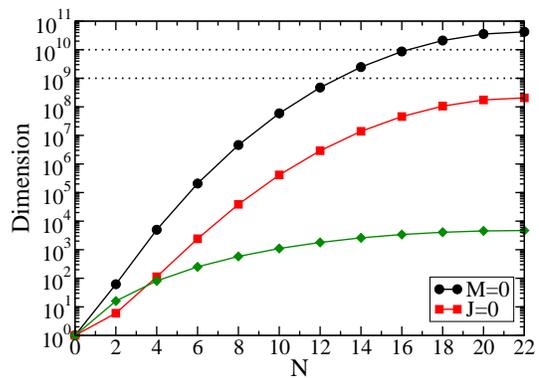}
\end{center}
\caption{
(Color Online) Dimensions of the $M^{\pi}=0^+$ (circle) and the corresponding $J^{\pi}=0^+$ (square) states in even-even Pb isotopes with $N<126$ as a function of valence neutron (hole) numbers. $M$, $J$ and $N$ denote the total magnetic quantum number, total spin and the number of valence neutrons, respectively. The green diamonds correspond to the total number of partitions (particle distributions).  The dashed lines correspond to the limit of present shell-model calculations. \label{dimension}
}
\end{figure}

In the present work we assume the doubly-magic $^{208}$Pb as the inert core.  Calculations are done in the hole-hole channel. 
The valence model space contains six orbitals between the magic numbers $N=82$ and 126, namely $2p_{1/2}$, $1f_{5/2}$,
$2p_{3/2}$, $0i_{13/2}$, $1f_{7/2}$ and $0h_{9/2}$. 
 In Fig. \ref{dimension} we plotted the $M$-scheme dimensions for the $M=0$ positive-parity states and the dimensions of the corresponding $I^{\pi}=0^+$ states in even-even Pb isotopes. The computational limit for contemporary shell-model calculations is around $10^{10}$ (in the $M$-scheme) for systems
with roughly the same numbers of protons and neutrons.
This is possible by applying the so-called factorization technique \cite{ant,John13}.
Systems with only identical particles are more difficult to treat numerically. In Refs. \cite{Qi12,Back11} we managed to do systematic shell-model calculations for Sn isotopes, for which largest system treated has the dimension around $10^9$.
 As for Pb isotopes in the present work,
the largest system we handled has a dimension $3.4\times 10^{9}$. This allows us to diagonalize in the full model space for all nuclei between $^{194-206}$Pb.

A common practice in full configuration interaction shell model calculations is to express the effective Hamiltonians in terms of single-particle energies and two-body matrix elements as
\begin{eqnarray}
\nonumber H&=&\sum_{\alpha}\varepsilon_{\alpha}{\hat N}_{\alpha} \\
&&+ \frac{1}{4}\sum_{\alpha\beta\delta\gamma JT}\langle j_{\alpha}j_{\beta}|V|j_{\gamma}j_{\delta}\rangle_{JT}A^{\dag}_{JT;j_{\alpha}j_{\beta}}A_{JT;j_{\delta}j_{\gamma}},
\end{eqnarray}
where $\alpha=\{nljt\}$ denote the single-particle orbitals and $\varepsilon_{\alpha}$ stand for the corresponding single-particle energies. $\hat{N}_{\alpha}=\sum_{j_z,t_z}a_{\alpha,j_z,t_z}^{\dag}a_{\alpha,j_z,t_z}$ is the particle number operator. $\langle j_{\alpha}j_{\beta}|V|j_{\gamma}j_{\delta}\rangle_{JT}$ are the two-body matrix elements coupled to good spin $J$ and isospin $T$. $A_{JT}$ ($A_{JT}^{\dag}$) is the fermion pair annihilation (creation) operator. 
The single-hole energies of above orbitals (relative to the $2p_{1/2}$ orbital) are taken from the experimental single-particle energies of the nucleus $^{207}$Pb as $\varepsilon_{2p_{1/2}}=0.0$ MeV, $\varepsilon_{1f_{5/2}}=0.57$ MeV,
$\varepsilon_{2p_{3/2}}=0.898$ MeV, $\varepsilon_{0i_{13/2}}=1.633$ MeV, $\varepsilon_{1f_{7/2}}=2.34$ MeV and $\varepsilon_{0h_{9/2}}=3.414$ MeV.

For the model space we have chosen the effective Hamiltonian such that it contains five single-particle energies relative to the $2p_{1/2}$ orbital and 353 $T=1$ two-body matrix elements. 
We take the interaction that has been continuously developed by the Stockholm
group in the past few decades \cite{Jan}. The mass dependence of the 
effective interaction is not considered in the present work, which is not expected to play any significant role for nuclei of concern.

The monopole interaction is defined as the angular-momentum-weighted average value of the diagonal matrix elements $\langle j_{\alpha}j_{\beta}|V|j_{\alpha}j_{\beta}\rangle_{JT}$ for a given set of $j_{\alpha}$, $j_{\beta}$ and $T$. For the chosen model space there are 21 $T=1$ (neutron-neutron) monopole terms. The strength of those monopole terms are given in Table \ref{ngp}. As can be seen from the table, the $T=1$ monopole interactions are small and are  mostly close to zero. This is consistent with shell-model calculations in light and medium-mass nuclei (see, e.g., Refs. \cite{hon09,jen95})
In Ref. \cite{Qi12} the monopole interaction for Sn isotopes was optimized by fitting to all low-lying states in Sn isotopes using a global optimization method. 
In the present work, they are determined by fitting to isotopes around $^{208}$Pb due to computation limitations.
We will neglect isospin below for simplicity since the systems we handle in the present work only contain valence neutron holes.

\begin{table}
  \centering
  \caption{The strengths of the monopole interactions for different orbitals as the average of the two-body matrix elements $\langle j_{\alpha}j_{\beta}|V|j_{\alpha}j_{\beta}\rangle_{JT}$.}\label{ngp}
\scriptsize
  \begin{tabular}{*{26}{ccccccc}}
  \hline  \hline
&$2p_{1/2}$& $1f_{5/2}$&
$2p_{3/2}$& $0i_{13/2}$& $1f_{7/2}$ & $0h_{9/2}$\\
\hline
$2p_{1/2}$&-0.0500&\\
 $1f_{5/2}$ &0.0504& 0.00833& \\
$2p_{3/2}$&0.00625& 0.0241& -0.0913&\\
 $0i_{13/2}$& 0.0394 & 0.0176&0.0822&-0.00357&\\
  $1f_{7/2}$ &0.0467 & 0.0141& 0.0149& 0.114&-0.00661&\\
  $0h_{9/2}$ & 0.0242 & 0.0886&0.0613&0.00020&0.0482&0.0923\\
  \hline  \hline
  \end{tabular}
\end{table}

The calculated total energy for a given isotope with $N$ valence neutron holes can be written as
\begin{equation}\label{esm}
E^{\rm tot} =C+N\varepsilon_{0}+ \langle\Psi_I|H|\Psi_I\rangle,
\end{equation}
where $\Psi_I$ is the calculated shell-model wave function and $I$ is the total angular momentum.  The constants $C$ and $\varepsilon_{0}$ denote the experimental (negative) binding energy of the core $^{208}$Pb and the one-neutron separation energy of the $2p_{1/2}$ state in $^{207}$Pb, respectively.
The excitation energy and wave function of a given state only depend on the shell model Hamiltonian $H$. One may rewrite the Hamiltonian as $H=H_m+H_M$ where $H_m$ and $H_M$ denote the (diagonal) monopole and Multipole Hamiltonians, respectively. The shell model energies can be written as
\begin{eqnarray}
\nonumber E^{\rm SM}&=&\langle\Psi_I|H|\Psi_I\rangle\\
\nonumber &=& \sum_{\alpha}\varepsilon_{\alpha}<\hat{N}_{\alpha}>+\sum_{\alpha\leq\beta}V_{m;{\alpha}\beta}\left<\frac{\hat{N}_{\alpha}(\hat{N}_{\beta}-\delta_{\alpha\beta})}{1+\delta_{\alpha\beta}}\right>\\
&&+\langle\Psi_I|H_M|\Psi_I\rangle.
\end{eqnarray}

All shell-model calculations are carried out within the $M$-scheme where states with $M=I$ are considered. Diagonalizations are done with a parallel shell model program that we developed \cite{Qi08} as well as the code KSHELL \cite{Shi13} with modifications.
The calculations are done on the supercomputers Beskow and Tegn\'er at PDC Center for High Performance Computing at the KTH Royal Institute of Technology in Stockholm, Sweden.  

\subsection{Monopole-based truncation}
Only $M$ ($J_z$) and $T_z$ are good quantum numbers in the M-scheme, leading
to a maximal dimension of the bases. It is also difficult to apply the variety of truncation algorithms since angular momentum is not explicitly conserved for a given basis \cite{Ro09,Kru2013,Jiao2014,Pap2005}. 
As a result,  it may become problematic if part of the bases within a given partition is removed from the model space. On the other hand, angular momentum conservation will not be a problem if one implements the truncation by considering a limited number of partitions and taking into account all  M-scheme bases within a given partition. A common approach is to apply the so-called n-particle-n-hole truncation by considering a limited number of particle-hole excitation across a presumed subshell. If several harmonic oscillator major shells are considered, the $n\hbar\omega$ (or $N^{max}$) truncation can be applied by limiting the total number of excitations crossing the major shells. However, such calculations do not consider the relative importance of the different configurations within a  particular particle-hole excitation. The convergence can be slow if there is no clear shell or subshell closure or if the single-particle structure can be significantly modified by the monopole interaction.

\begin{figure}  
\begin{center}
\includegraphics[width=0.4\textwidth]{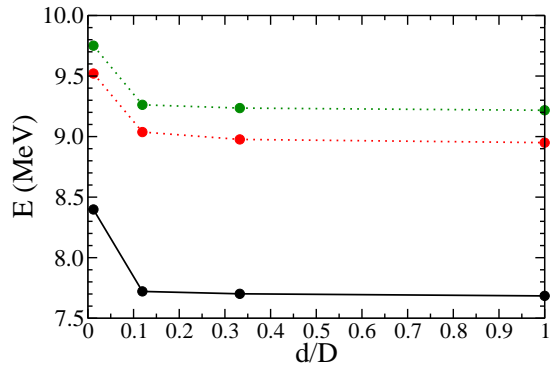}
\end{center}
\caption{\label{fig:c194}  (color online)  Convergence of the shell-model energies for the lowest three states in nucleus $^{194}$Pb  as a function of the fraction of the M-scheme bases consider. $D$ is the total number of bases while $d$ denotes the number of bases considered in the truncated shell model calculations. 
}
\end{figure}

Alternatively, one may consider an importance truncation based on the total monopole energy by considering the multipole Hamiltonian as a perturbation.    
One can evaluate the total monopole energy of a given partition as
\begin{eqnarray}
\nonumber E^{\rm m}_P&=& \sum_{\alpha}\varepsilon_{\alpha}{N}_{P;\alpha}+\sum_{\alpha\leq\beta}V_{m;{\alpha}\beta}\frac{N_{P;\alpha}(N_{P;\beta}-\delta_{\alpha\beta})}{1+\delta_{\alpha\beta}},
\end{eqnarray}
where $N_{P;\alpha}$ denotes the particle distributions within a given partition $P$. One can order all partitions according to the monopole energy $E^{\rm m}_P$ and consider the lowest ones for a given truncation calculation. Moreover, it is expected that the pairing correlation should play a significant role governing the structure of the lowest-lying states of the semi-magic Pb isotopes. 
Taking those facts into account, we have done truncation calculations by considering the relative importance as defined by the monopole Hamiltonian and monopole plus diagonal pairing Hamiltonian for two isotopes $^{200,194}$Pb. In both cases, convergence can be reached with a small portion of the full M-scheme wave functions. Moreover, for the lowest-lying states, the latter calculation converges noticeably faster.
In Fig. \ref{fig:c194} we plotted the results for the lowest three positive-parity states in the isotope $^{194}$Pb. Convergence is practically reached at $d/D\sim 0.1$, i.e.,  by considering only $10\%$ of the total M-scheme bases. The result for $^{200}$Pb is given in the supplementary material.

\subsection{The generalized seniority}
Generalized seniority has long been introduced \cite{Talmi_1971,
Shlomo_1972, Allaart_1988, Gambhir_1969} as a truncation scheme for the nuclear shell model and has been applied to explain the structure and transition properties of many spherical nuclei including the Sn isotopic chain \cite{Mor11}. In this work, calculations are performed by considering states with generalized seniority up to $S=6$ (six unpaired neutron holes), where the coherent pair structure is determined by minimizing the average energy of the fully-paired ($S=0$) state.
High generalized seniority could be reached by pre-calculating the ``many-pair density matrix'' characterizing the pair condensate as we proposed recently (see Ref. \cite{Jia_2015} for technical details).

\subsection{The nucleon-pair approximation of the shell model}
The nucleon-pair approximation (NPA) is a pair-truncation scheme for the nuclear shell model \cite{NPA1,NPA2,NPA3}. The building blocks of the NPA are collective nucleon pairs with given spin and parity. For $2N$ valence nucleons outside a doubly-magic nucleus, the configuration space is constructed by such pairs coupled successively as
 \begin{eqnarray} \label{NPA1}
 {A^{(J_N)}}^{\dagger} (r_1 \cdots r_N,J_1 \cdots J_N) \nonumber \\
  \quad\quad\quad\quad \equiv [ \cdots ( ( {{A}^{(r_{1})}}^{\dag}\times {{A}^{(r_{2})}}^{\dag} )^{(J_2)} \times {{A}^{(r_{3})}}^{\dag})^{(J_3)}\nonumber \\
  \times \cdots \times {{A}^{(r_{N})}}^{\dag} ]^{(J_N)} ,\nonumber
 \end{eqnarray}
where $(r_i)$ is short for ($J_{r_i}$ and $\pi_{r_i}$), and $(J_i)$ short for ($J_i$ and $\pi_i$).
${{A}^{(r_{i})}}^{\dag}=\sum_{ab}y( ab r_i )({a}^{\dagger} \times {b}^{\dagger})^{(r_i)}$ denotes a collective nucleon pair with spin $J_{r_i}$ and parity $\pi_{r_i}$; ${a}^{\dagger}$ (${b}^{\dagger}$) is the creation operator of a nucleon in the single-particle orbit $a$ ($b$); $y( ab r_i )$ is called the pair structure coefficient.

\section{Results}
\begin{figure}  
\begin{center}
\includegraphics[width=0.49\textwidth]{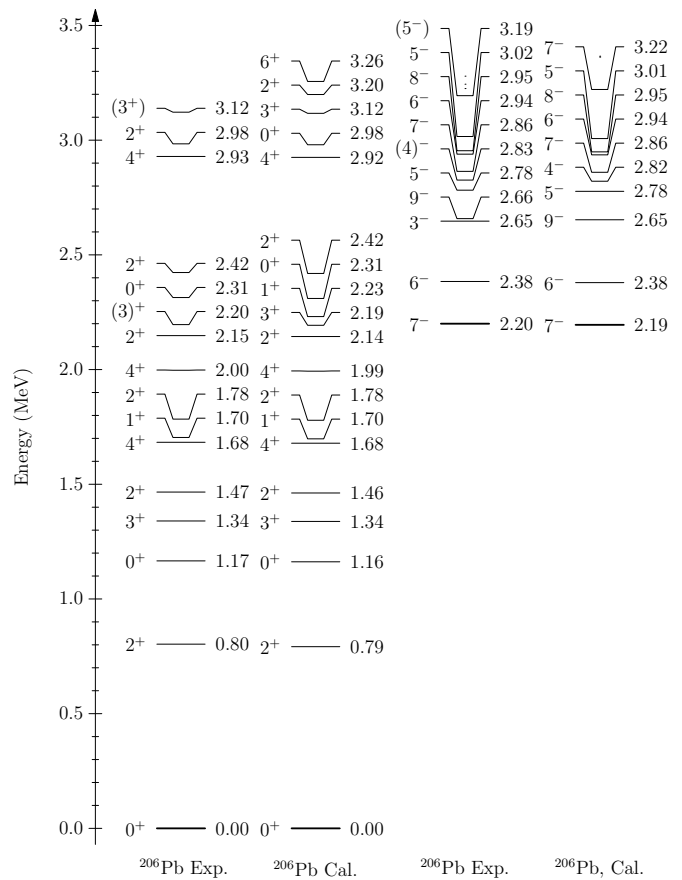}
\end{center}
\caption{\label{fig:pb206} Experimental data and shell-model calculations on the low-lying spectrum of $^{206}$Pb.
}
\end{figure}

\begin{figure}  
\begin{center}
\includegraphics[width=0.49\textwidth]{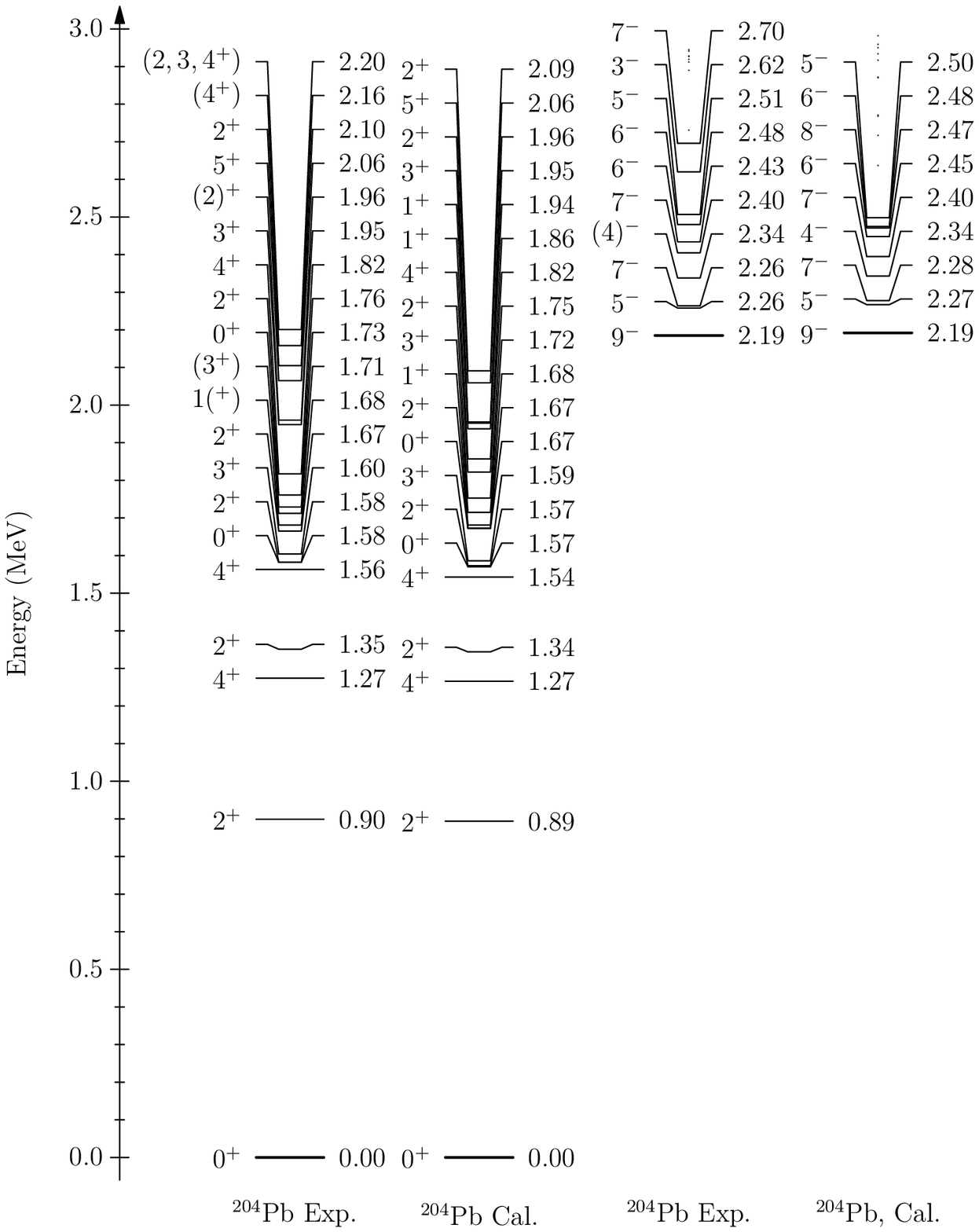}
\end{center}
\caption{\label{fig:204} Experimental and shell-model calculated low-lying spectra of $^{204}$Pb.
}
\end{figure}

\begin{figure}  
\begin{center}
\includegraphics[width=0.49\textwidth]{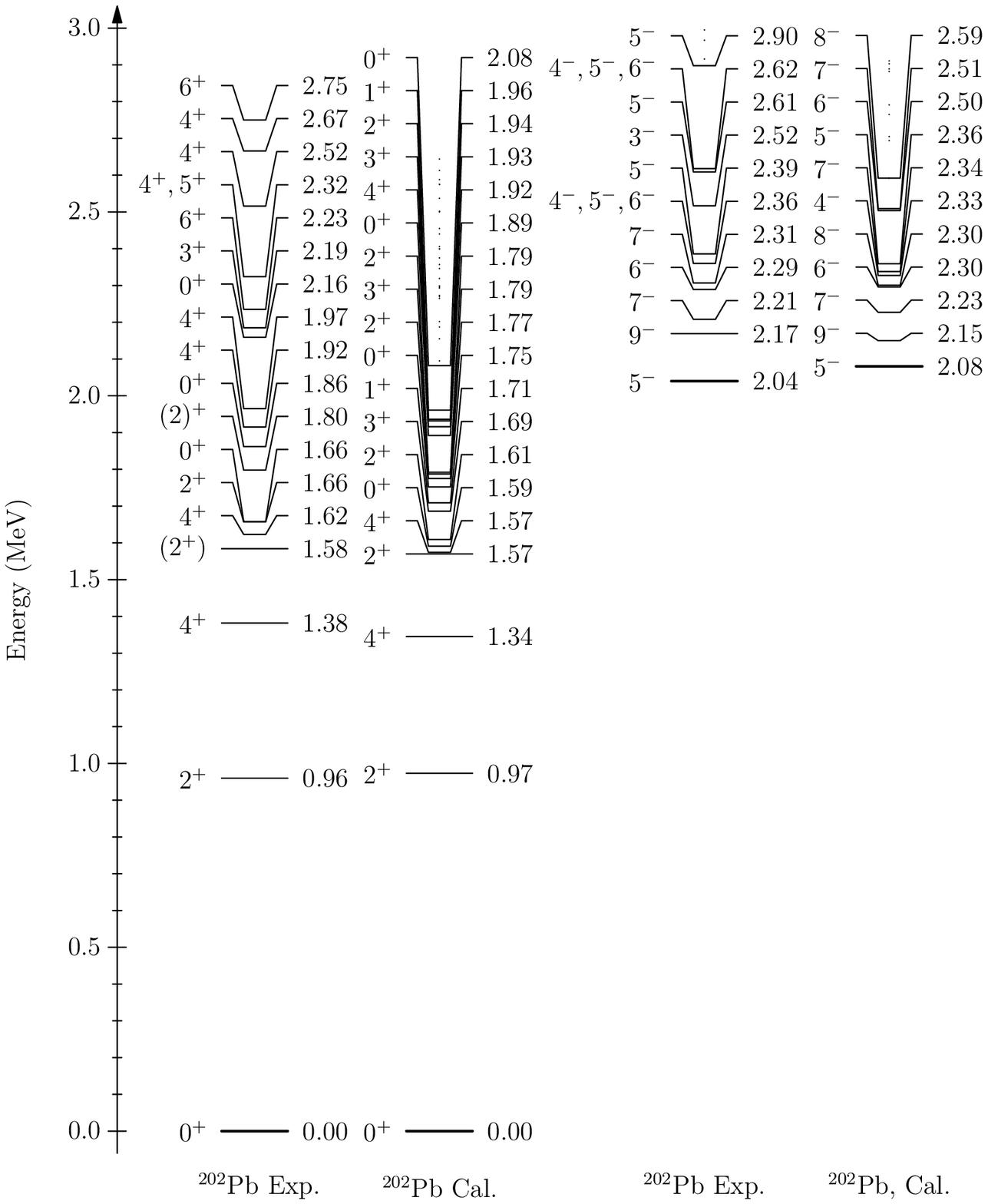}
\end{center}
\caption{\label{fig:202} Experimental and shell-model calculated low-lying spectra of $^{202}$Pb.
}
\end{figure}

\begin{figure}  
\begin{center}
\includegraphics[width=0.49\textwidth]{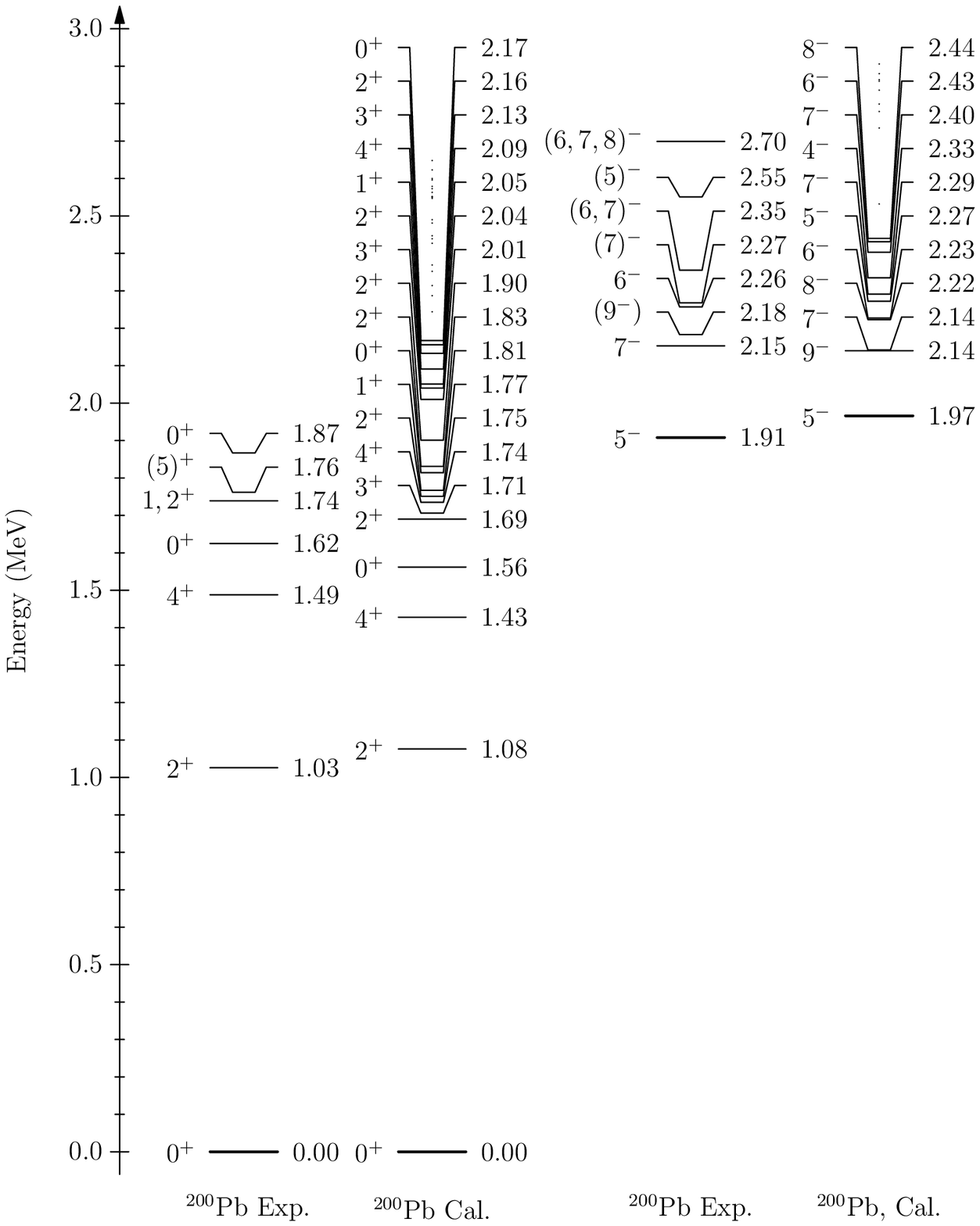}
\end{center}
\caption{\label{fig:200} Experimental and shell-model calculated low-lying spectra of $^{200}$Pb.
}
\end{figure}

\begin{figure}  
\begin{center}
\includegraphics[width=0.49\textwidth]{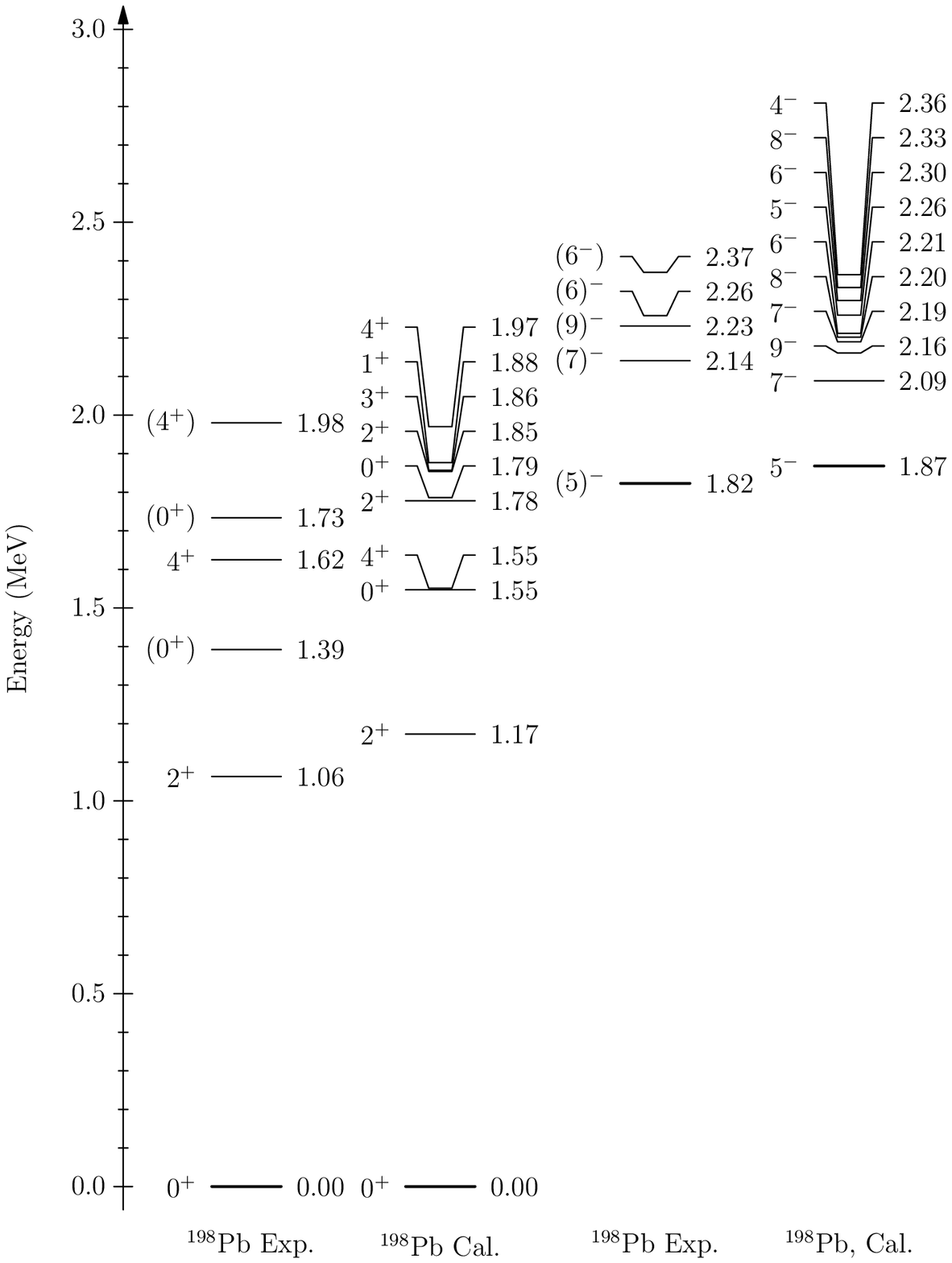}
\end{center}
\caption{\label{fig:198}  Experimental and shell-model calculated low-lying spectra of $^{198}$Pb.
}
\end{figure}

\begin{figure}  
\begin{center}
\includegraphics[width=0.49\textwidth]{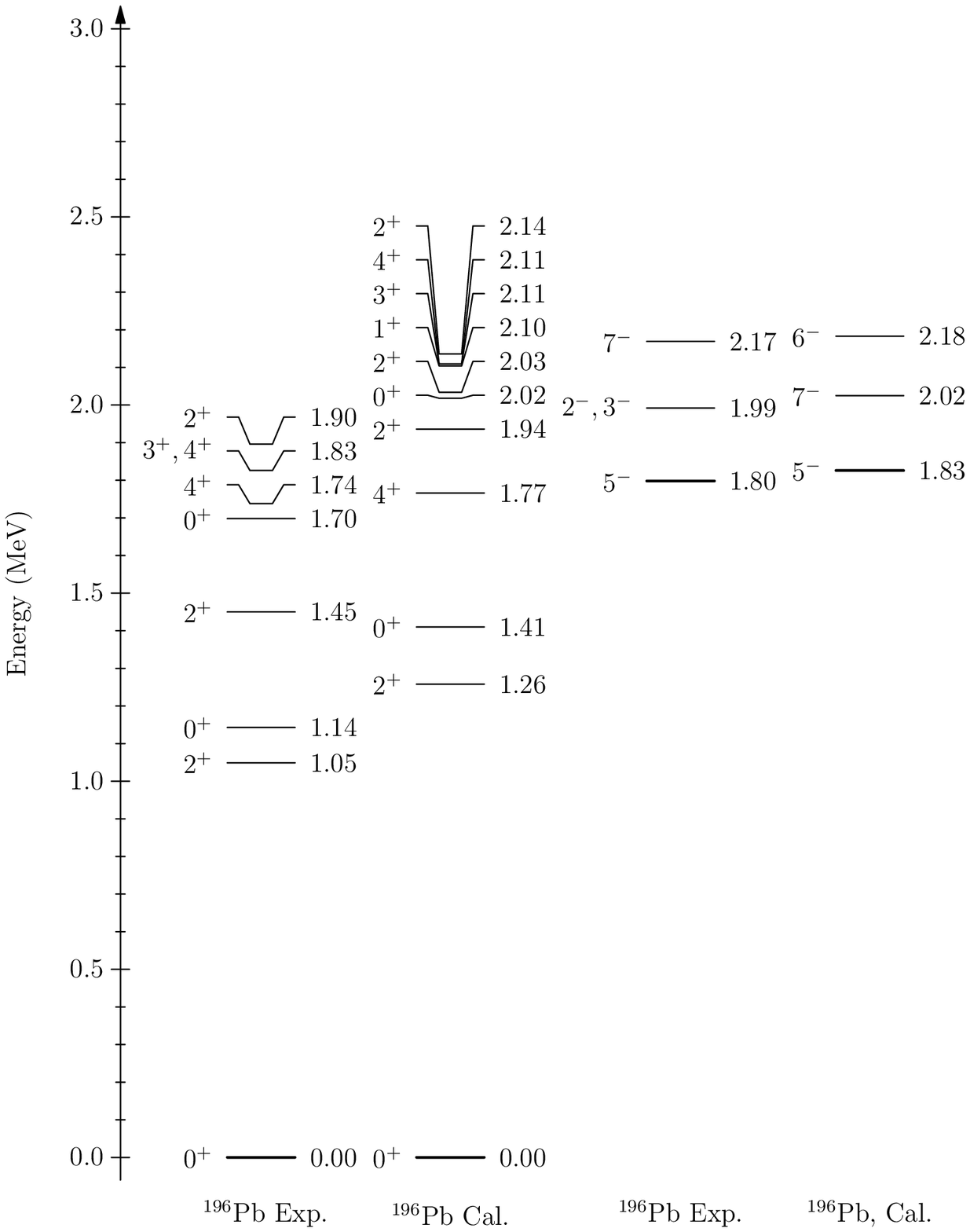}
\end{center}
\caption{\label{fig:196} Experimental and shell-model calculated low-lying spectra of $^{196}$Pb.
}
\end{figure}

Shell-model calculations with different interactions have been reported for nuclei $^{204,206}$Pb in Refs. \cite{Cor98,Sha12,Mc75,wang90}.
Our calculations for these nuclei are given in Figs. \ref{fig:pb206} and \ref{fig:204} together with experimental data from Ref. \cite{nudat}. An excellent agreement between theory and experiment is obtained for all even isotopes down to $^{196}$Pb. The results for $^{196-202}$Pb are shown in Figs. \ref{fig:202}, \ref{fig:200}, \ref{fig:198} and \ref{fig:196}. We also compared our results with calculations using the realistic Bonn potential in Ref. \cite{Cor98}. In the latter cases, the spectra for the two even-even systems are both compressed in comparison with experimental data and the energies of the excited states are systematically underestimated.
The gaps between the $7^-$ and $6^-$ state in $^{206}$Pb  and those between different $0^+$ and $2^+$ states were also underestimated.   
 In our calculation,
the lowest four positive-parity states including $0^+_2$  are dominated by the coupling of neutron hole pair in the orbitals $2p_{1/2}$ and $1f_{5/2}$ where the largest spin corresponds to $3^+$. For states above that one, the coupling to $2p_{3/2}$ becomes important.
The $7^-$ and $6^-$ states are due to the coupling between $2p_{1/2}$ and $0i_{13/2}$. The collective core-excited $3^-$ state is beyond the scope of the present model space. The other negative-parity states are mainly composed by $1f_{5/2}\otimes 0i_{13/2}$. The states in the nucleus $^{204}$Pb show a similar structure.
The $7^-$ and $6^-$ states get disfavored in energy in relation to the enlarged occupation of the $2p_{1/2}$ orbital.
As for the nucleus $^{206}$Pb, there are six $0^+$ states in total within the model space. Four of them are calculated to be lower than 3 MeV.
In $^{204}$Pb, the $9^-$ state becomes the lowest lying negative parity state. It is dominated by the configuration  $(1f_{5/2}\otimes0i_{13/2})^{9^-}$. On the other hand, a mixture between $(1f_{5/2}\otimes0i_{13/2})^{5^-}$ and $(2p_{3/2}\otimes0i_{13/2})^{5^-}$ are seen for the lowest $5^-$ state. The $5^-$ state becomes the lowest negative parity states in the isotopes $^{188-202}$Pb.
The low-lying spectrum of $^{202}$Pb shows a large similarity with those of $^{200,204}$Pb. This is because the wave functions in all these states contain a large contribution from the coupling of particles in $1f_{5/2}$.

\begin{figure}  
\begin{center}
\includegraphics[width=0.4\textwidth]{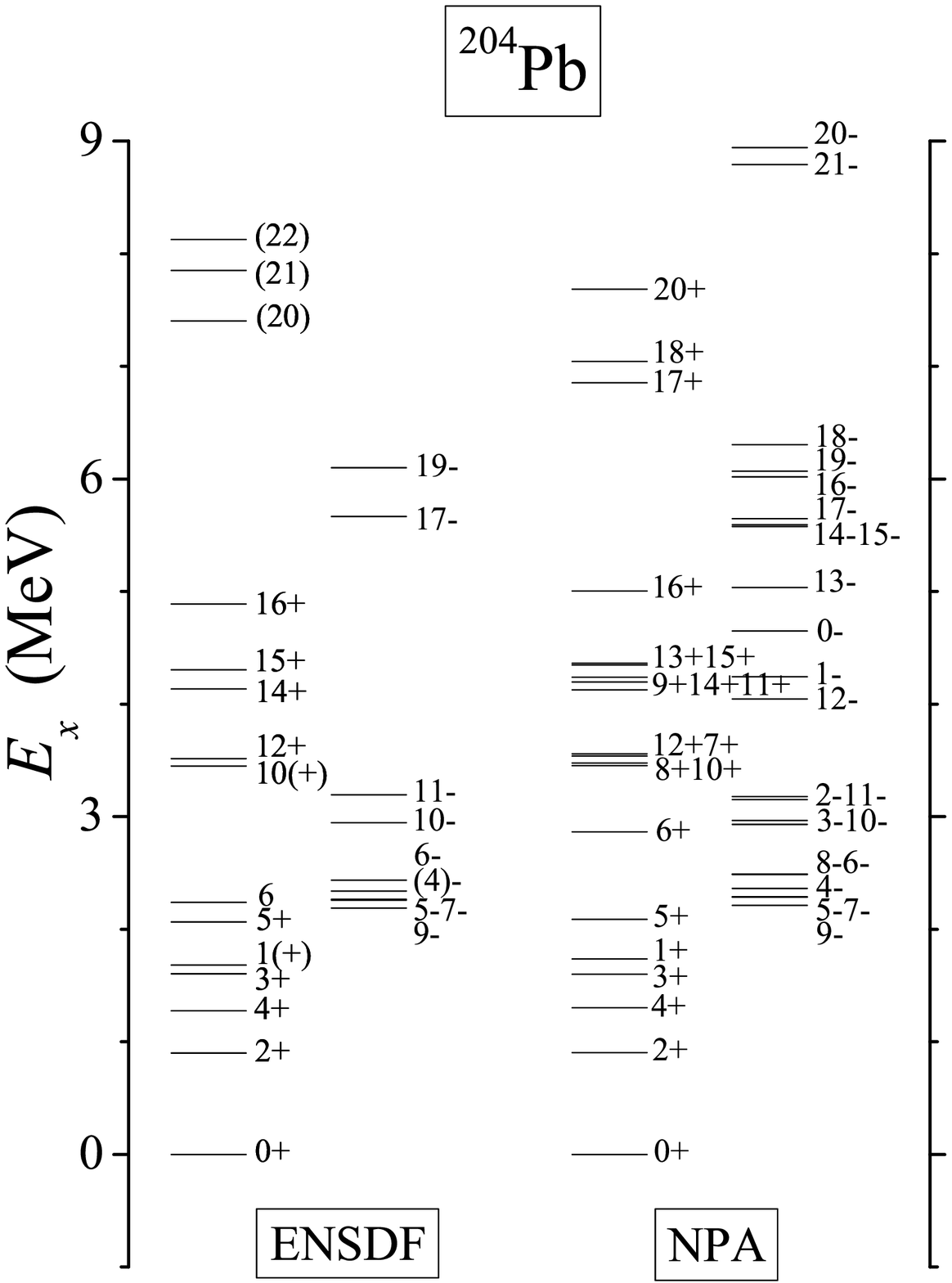}
\end{center}
\caption{\label{fig:204npa} Experimental and NPA calculated low-lying spectra of $^{204}$Pb.
}
\end{figure}

 For calculations with the realistic interaction as employed in the present work, our NPA calculation is currently limited to systems with upto 4 pairs (a maximum of 8 pairs if the schematic P+QQ Hamiltonian is used). In the present work we choose the NPA configuration space constructed by a few important nucleon pairs for $^{204}$Pb, $^{202}$Pb, and $^{200}$Pb. For the yrast states of $^{204}$Pb, $^{202}$Pb, and $^{200}$Pb, the level energies and electromagnetic properties obtained by the NPA are very close to those obtained by the shell model or the generalized seniority scheme. The results are given in Figs. \ref{fig:204npa}, \ref{fig:202npa} and \ref{fig:200npa}. A detailed analysis of the NPA wave function is given in the supplementary material.

\begin{figure}  
\begin{center}
\includegraphics[width=0.4\textwidth]{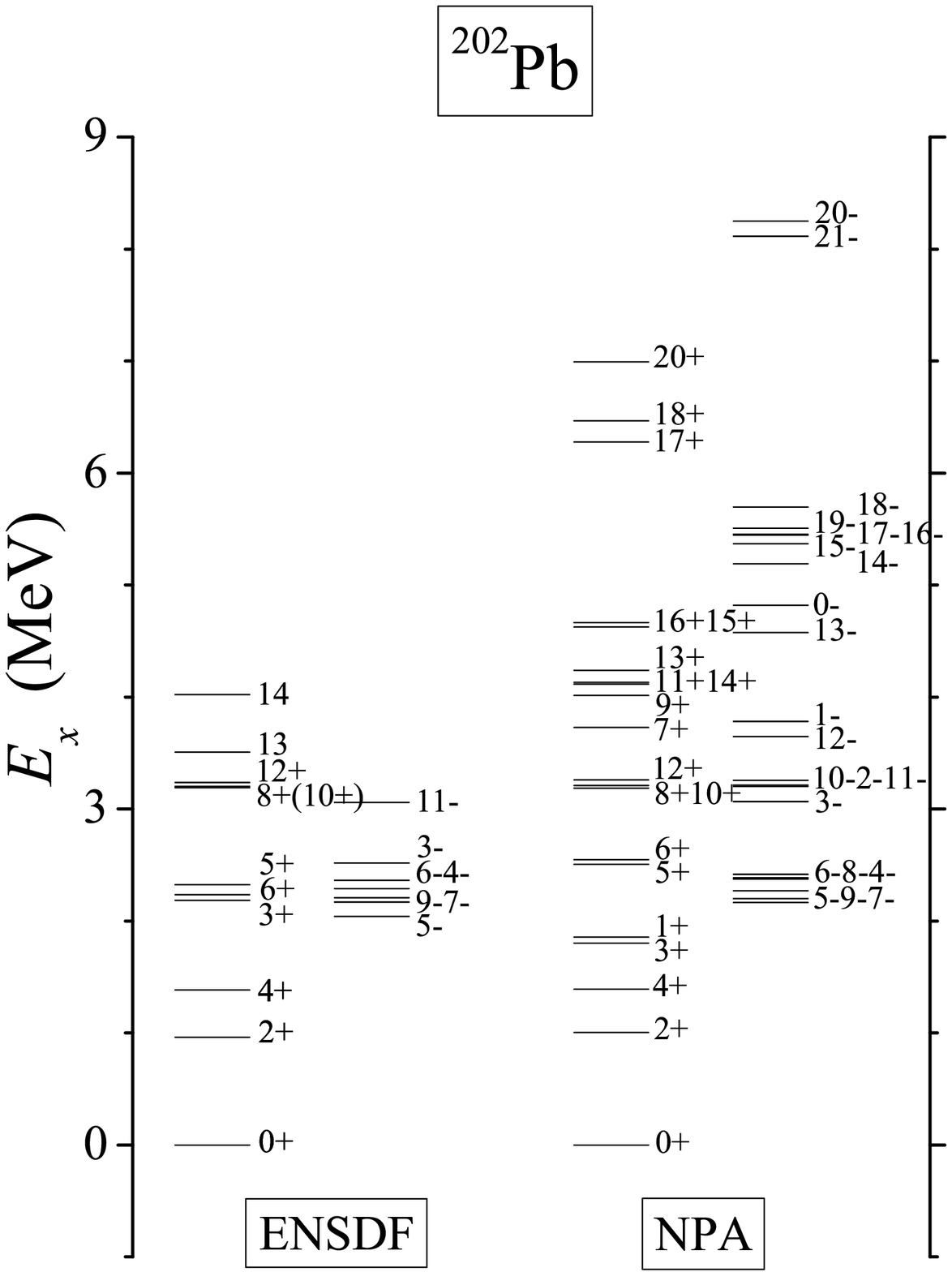}
\end{center}
\caption{\label{fig:202npa} Experimental and NPA calculated low-lying spectra of $^{202}$Pb.
}
\end{figure}

\begin{figure}  
\begin{center}
\includegraphics[width=0.4\textwidth]{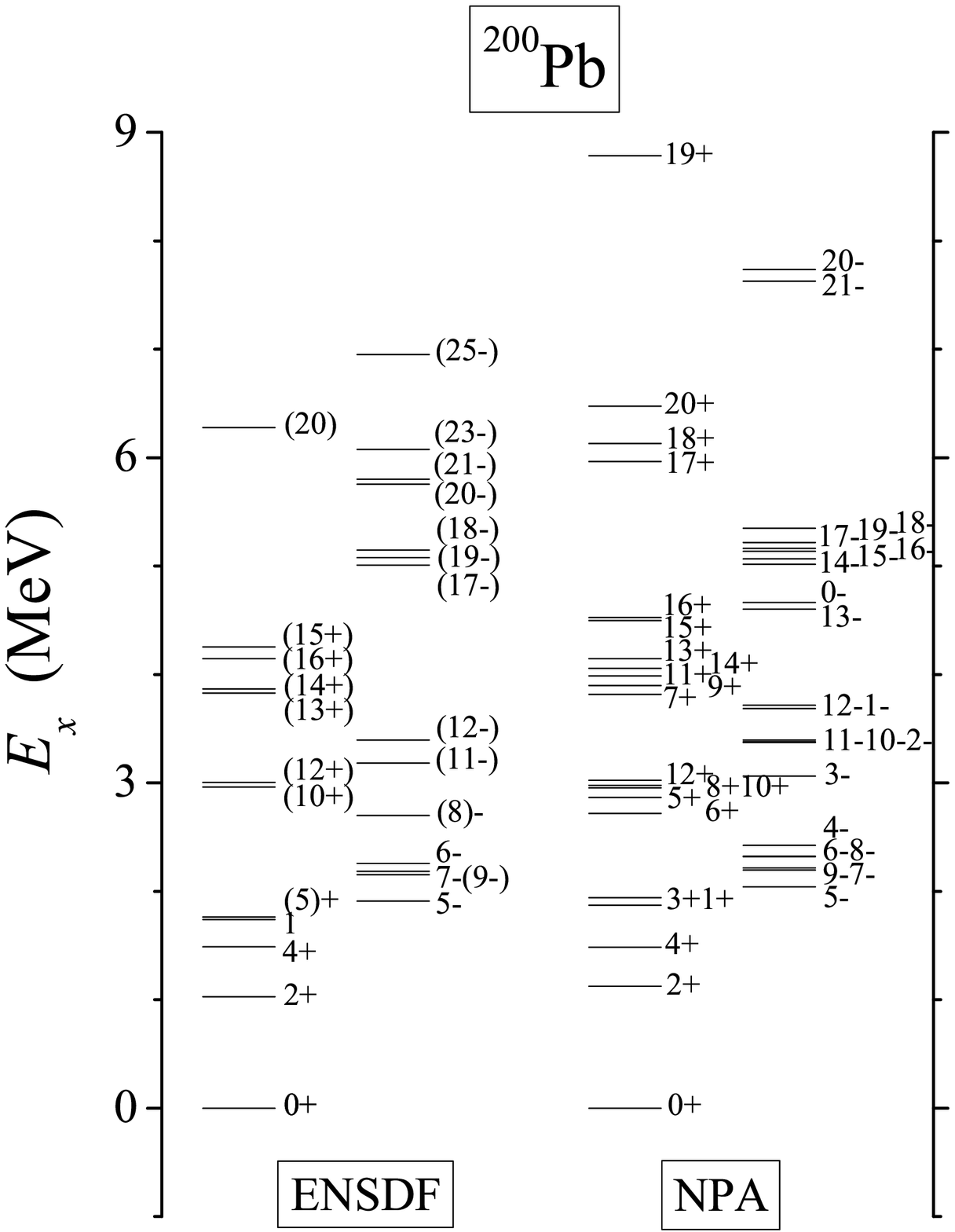}
\end{center}
\caption{\label{fig:200npa} Experimental and NPA calculated low-lying spectra of $^{200}$Pb.
}
\end{figure}

We have also done systematic calculations for the low-lying spectra of all odd Pb isotopes $^{195-205}$Pb. An overall good agreement is also obtained from which a one to one correspondence between theory and experiment can be easily identified. In $^{205}$Pb, there are two low-spin negative-parity states observed with the energies 0.803 and 0.995 MeV. There are two calculated states in this region with the spin-parity (energy) of $1/2^-$ (0.891 MeV) and $3/2^-$ (1.086 MeV), respectively. The first positive-parity state is calculated to be $13/2^+$ at  1.107 MeV, which can be compared to the experimental result of 1.014 MeV.
The first $5/2^-$ and $1/2^-$ states in $^{205}$Pb are nearly degenerate, for which the splitting increases to 126.5 keV in $^{203}$Pb. The calculated gap is 143 keV.
The excitation energy of the $13/2^+$ state reduces to 825 keV for which the calculated value is 849 keV. The second $13/2^+$ state is calculated to be at 1.657 MeV.
The excitation energy of the $13/2^+$ state reduces further as neutron numbers decrease. In $^{197}$Pb, the excitation energy is only 319 keV in comparison to the calculated value of 362 keV. From above good agreement one may safely state that the present shell-model calculations are able to correctly reproduce the evolution of the single-particle structure of Pb isotopes and the monopole interactions are mostly under control.

\begin{figure}  
\includegraphics[width=0.35\textwidth]{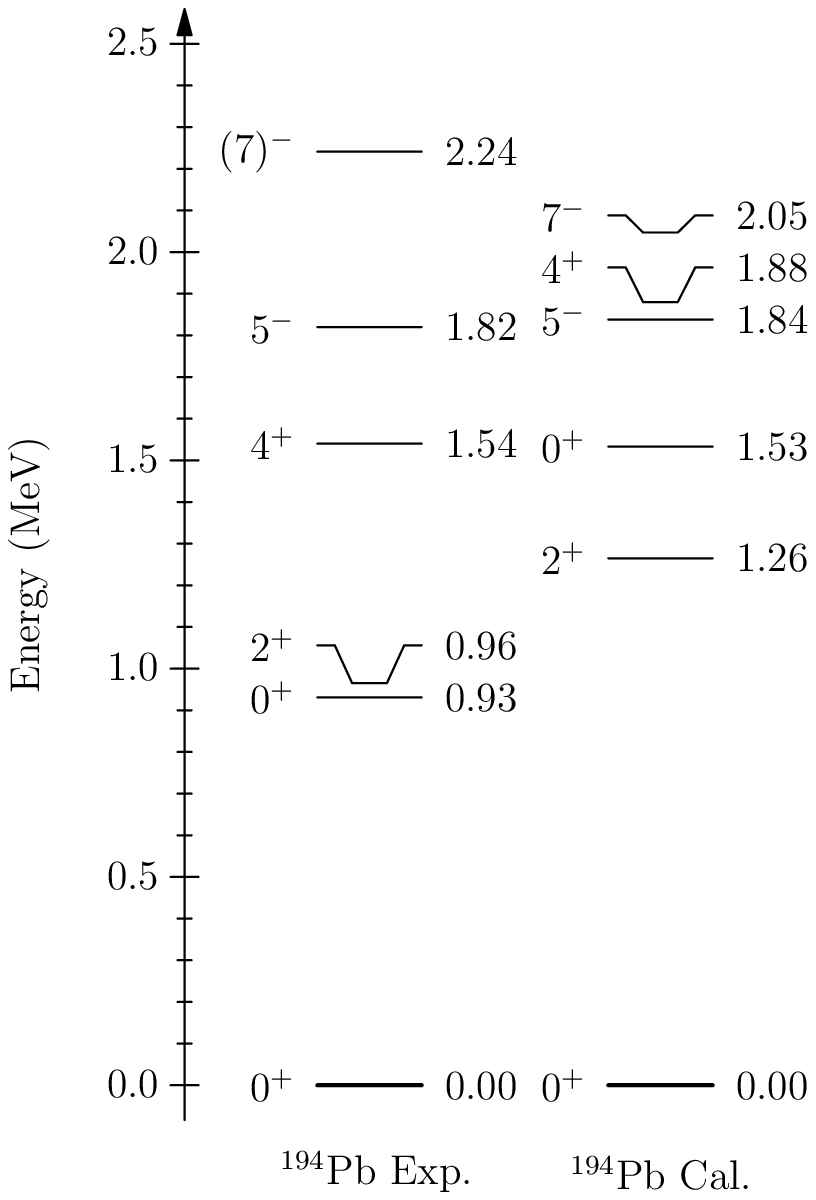}
\caption{\label{fig:194} Experimental and shell-model calculated low-lying spectra of $^{194}$Pb.
}
\end{figure}

The nucleus $^{194}$Pb is the lightest system we can do the full shell-model calculation. The results are shown in Fig. \ref{fig:194}. For this nucleus, the shell-model calculation overestimates the excitation energies of the first $2^+$ state by 300 keV and first excited $0^+$ state by 600 keV. It also fails to reproduce the inversion between the two excited states.

\begin{figure}  
\begin{center}
\includegraphics[width=0.45\textwidth]{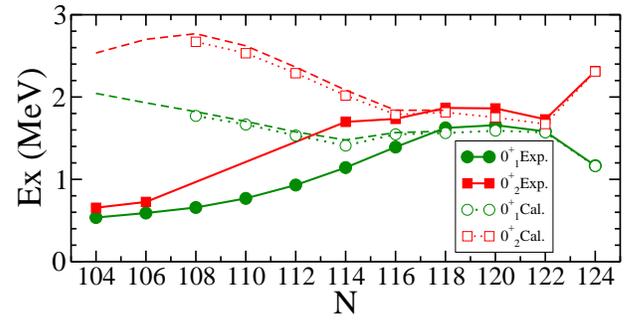}
\end{center}
\caption{\label{fig:0+}  (color online) Left: Experimental \cite{nudat} (solid symbols) and shell-model calculated (open symbols) excitation energies for the first-excited (circle) and second-excited (square) $0^+$ states in Pb isotopes as a function of neutron number. The dashed lines correspond to calculations with the generalized seniority model with the truncation $v=6$.
}
\end{figure}

Our shell-model calculations can reproduce well the excitation energies of the low-lying $0^+$ states in isotopes $^{198-206}$Pb. The evolution of the excitation energies of the first two excited $0^+$ states are summarized in Fig. \ref{fig:0+}. As can be seen from the figure, large deviations are only seen for the excited $0^+$ states in nuclei lighter than $^{196}$Pb.
In those lighter Pb isotopes, the excitation energy of the second $0^+$ state decreases rapidly with decreasing
neutron number. It even becomes the first excited state in $^{184-194}$Pb. 
Within a shell-model context, those low-lying $0^+$ states may be interpreted as coexisting deformed states which are induced by proton pair excitations across the $Z=82$ shell gap \cite{Hey87}. The energy of those core-excited configurations get more favored in mid-shell Pb isotopes in relation to the stronger neutron-proton correlation in those nuclei. 
The existence of several competing minima in those  neutron-deficient nuclei can already be n clearly see in potential energy surface calculations with deformed Woods-Saxon potentials \cite{Ben89,Ben93}.
Self-consistent mean-field calculations were also done \cite{Smi03,Lib99,Nik02,Yos94}. In particular,
beyond mean field calculations for the spectroscopy were done in Refs. \cite{Nom12,Yao13,Egi,Rod04,Fra,Hell08,Dug03,Bender04,Lib99}.
In Ref. \cite{Egi}, those coexisting states were also explained in terms of neutron correlations. In the future, it could be very interesting to explore this problem from two perspectives: One is the possible mixing effect between the spherical neutron-neutron correlation and the proton particle–hole excitations in those nuclei. 
On the other hand, if one assumes that those low-lying collective states are pure intruder states like the $3^-$ state in $^{206}$Pb, it may also be interesting to measure those excited spherical $0^+$ states as predicted Fig. \ref{fig:0+} by the shell model.

\begin{figure}  
\begin{center}
\includegraphics[width=0.45\textwidth]{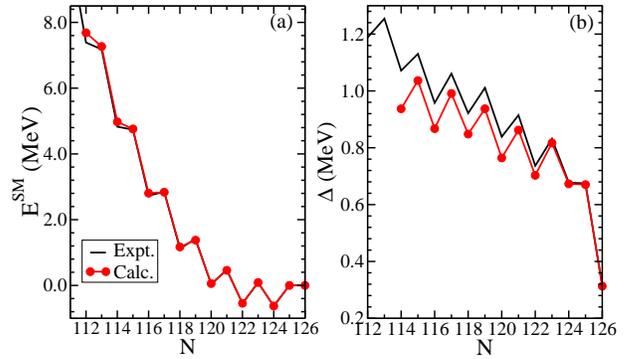}
\end{center}
\caption{\label{fig:delta}  (color online) (a): Experimental \cite{AM12} and  shell-model calculated shell-model correlation energies (Eq. (\ref{esm})) as a function of neutron number; (b): The empirical pairing gaps as extracted according to Eq. (\ref{dexp}).
}
\end{figure}

The present shell-model calculation also shows a good agreement with experiments for the total binding energies of both even and odd Pb isotopes. In Fig. \ref{fig:delta} we plotted the shell-model correlation energies, $E^{SM}$, and those extracted from experimental data according to Eq. (\ref{esm}). 
Those energies are defined in the hole-hole channel relative to the assumed core $^{208}$Pb. A more positive value would indicate less binding energy for a given state.
For nuclei heavier than $^{196}$Pb, the difference between theory and experiment is less than 100 keV. The largest deviation appears in the case of $^{194}$Pb for which the calculation overestimate $E^{SM}$ by 300 keV (which means that the total binding energy is underestimated).

The empirical pairing gaps can 
be extracted from the experimental and calculated binding energies by using the simple three-point formula as (see, e.g., Refs. \cite{And13,PhysRevC.91.024305,Qi12} and references therein)
\begin{equation}\label{dexp}
\Delta_n(N)=\frac{(-1)^{(N+1)}}{2}\left[E(N)+E(N-2)-2E(N-1)\right].
\end{equation}
These gaps may provide invaluable information on the two-neutron as well as $\alpha$ clustering in the nuclei involved. The results extracted from experimental and calculated binding energies for Pb isotopes
are shown as a function of the neutron number in the right panel of Fig. \ref{fig:delta}. As can be seen from the figure, the overall agreement between experiments and calculations on the pairing gaps are quite satisfactory. Noticeable differences are only seen for mid-shell nuclei $^{196-198}$Pb for which the calculation underestimates the experimental data by around 100 keV. The deviation is mainly related to the relatively large difference between experimental and calculated binding energies of the even nuclei $^{194,196}$Pb. This indicates that a further enhancement of the relevant $J=0$ pairing matrix elements, in particular that of $i_{13/2}$, may be necessary.

\section{Electromagnetic moments}
\begin{figure}
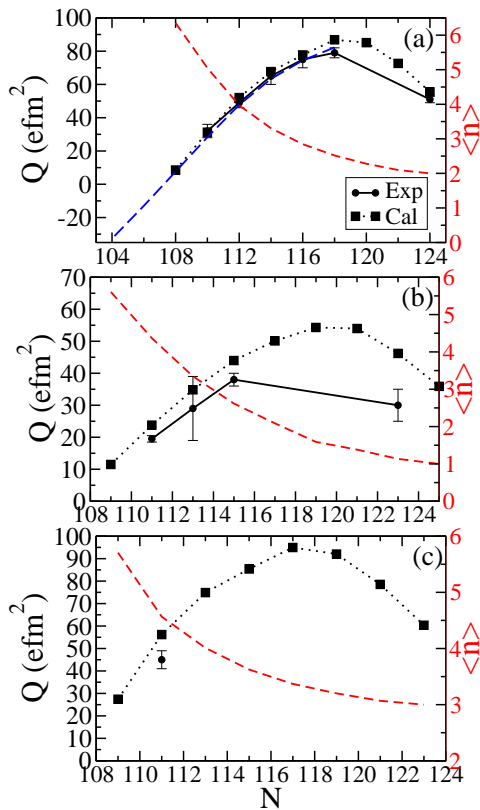
  
\begin{center}
\includegraphics[width=0.35\textwidth]{12+.eps}
\includegraphics[width=0.34\textwidth]{13+.eps}
\includegraphics[width=0.35\textwidth]{33+.eps}

\end{center}
\caption{\label{fig:12+}  (color online) Experimental (circle) \cite{Stone05} and shell-model calculated (square) quadrupole moments for the $12^+$ (a), $13/2+$ (b) and $33/2^+$ (c) states in even and odd Pb isotopes. The dashed lines correspond to the average number of neutron holes in the $i_{13/2}$ orbital. The blue long dashed line in the upper panel is the predictions by the generalized seniority calculations with $v=6$.
}
\end{figure}

There has been a long history measuring the quadrupole moments and magnetic moments of the isomeric $12^+$, $13/2^+$ and $33/2^+$ states in Pb isotopes \cite{zyw81,Nak72,Young75,Sten83,Sten85,Ruv86,Lin76,Blo93,Ion04,Ion07,Kmi10,Rou77}.
The $I^{\pi}=12^+$ isomers in the even-even Pb isotopes have been interpreted as $(vi_{13/2})^{-2}$ quasi-particle states. With decreasing neutron number, the single-hole character of the states was speculated to be influenced by the particle excitations from the core and become deformed. 
In the upper panel of Fig. \ref{fig:12+} we have  done full shell model calculations for the static quadrupole moments of the  $12^+$ states in even $^{194-206}$Pb isotopes and compared with available experimental data taken from Ref. \cite{Stone05}. Importance-truncation calculations are also done for the isotopes $^{190,192}$Pb. 
In this and all following calculations, we have simply taken $e_n=-1.0e$ for the effective charge of the neutron hole. There were indications that the effective charge may show a smaller absolute value with $e_n\sim-0.95e$ but it will not affect the general trends. We have also plotted in the upper panel of Fig. \ref{fig:12+} the results from generalized seniority calculations for isotopes $^{186-200}$Pb with the truncation $v=6$.
As can be seen from the panel, good agreement between theory and experiment is obtained for all available data. In the panel we also plotted the calculated average number of neutron holes in the orbital $i_{13/2}$. This quantity, as discussed in Refs. \cite{Bro92,Ney03}, can be very helpful in our understanding of the general trend of the quadrupole moments for those states. Indeed, the calculated quadrupole moments are dominated by the contribution from that orbital. As the occupancy increases, the quadrupole moments follow a linear decreasing trend and eventually vanish around half-filling. A similar correlation is also seen in the $10^+$ states in Sn isotopes \cite{Bro92,Qi12}.

In the middle panel of Fig. \ref{fig:12+} we plotted shell model calculations for the static quadrupole moments of the  $13/2^+$ states in odd $^{191-207}$Pb isotopes and compared with available experimental data.
Those $13/2^+$ isomers can be simply described as one-quasiparticle states in $vi_{13/2}$. The quadrupole moments in the lighter Pb isotopes follow a linear trend similar to that of the $12^+$ states, which is also related to the gradual occupancy of the $i_{13/2}$ orbital.
There is only one datum available for the heavier Pb isotopes, which seems overestimated by the present calculations. Further measurement may be necessary to clarify the issue.

In the lower panel of Fig.  \ref{fig:12+} we plotted shell model calculations for the static quadrupole moments of the  $33/2^+$ states which are dominated by the three quasi-particle configuration $(vi_{13/2})^{-3}$.
The calculations agree well with the only data available and follow a similar trend as the quadrupole moments of the  $12^+$ and $13/2^+$ states.

\begin{figure}  
\begin{center}
\includegraphics[width=0.4\textwidth]{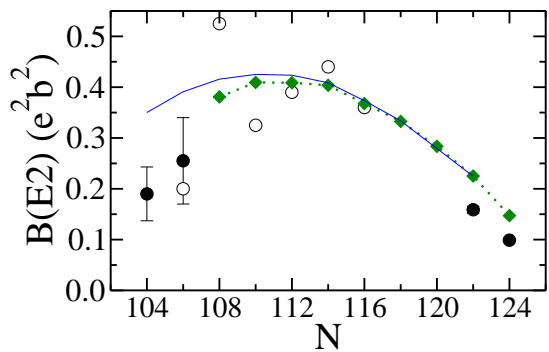}
\end{center}
\caption{\label{fig:be2}  (color online) Experimental \cite{Pri15} (circle) and shell-model calculated (diamond) B(E2; $0_1^+\rightarrow 2_1^+$) values for even-even Pb isotopes. The blue line corresponds to predictions by the generalized seniority model. The open symbols are preliminary results from Ref. \cite{Pak15}.
}
\end{figure}

The excitation energies of the first $2^+$ states in Sn isotopes between $^{102}$Sn and $^{130}$Sn are  established to possess an almost constant value. This is understood from the simple perspective of generalized seniority scheme~\cite{Mor11,Talmi_1971}.
As for Pb isotopes, the excitation energies of the first $2^+$ isotopes show a rather weak parabolic behavior, as can be seen in Fig. 1 in Ref. \cite{Haas86}.
Furthermore, the energy of the first excited $2^+$ state also gets systematically lowered for isotopes lighter than $^{196}$Pb. This trend is not reproduced by the calculation.
This indicates that the neglected deformed (or particle-hole) configurations may become important in those $2^+$ states. The measurement of the E2 transition between those states and the ground states are important in determining the structure of those states. 
Quasi-particle random phase approximation calculations for the B(E2) values of $^{204-210}$Pb were done in Ref. \cite{Ans05}.
In Fig. \ref{fig:be2} we have plotted our shell-model and generalized seniority truncation calculations for the B(E2) values of the transitions $0_1^+\rightarrow 2_1^+$. The available results for $^{186,188}$Pb seem being overestimated by our calculations, which may be related to the neglect of intruder configuration in our calculation.  More precise measurements may be necessary to clarify the discrepancy. Moreover, a slight difference is also seen between our shell-model and generalized seniority truncation calculations for the E2 transitions of $^{190-194}$Pb, for which the description may require the mixture of states with even higher seniority.

\section{Summary}
To summarize, the development of configuration interaction shell model algorithms makes it possible to study systematically of a long chain of isotopes on the same footing, which can provide a good testing ground for the shell model as well as the monopole channel of the two-body residual interaction. In the present work, we have carried out large-scale shell model calculations  to study
the structure properties of Pb isotopes with $N<126$. The effective interaction is optimized to reproduce the low-lying spectra of Pb isotopes close to $N=126$. It shows very good extrapolation properties. Both the ground state binding energies and excitation energies of low-lying states of the odd and even Pb isotopes can be reproduced very well. Larger deviations are only seen for the excited $0^+$ states in neutron-deficient Pb isotopes which are expected to be deformed states being dominated by intruder configuration. Our shell-model results also agree well with the high generalized seniority (with seniority quantum number up to $v=6$) and nucleon-pair-approximation truncation calculations.
Systematic calculations on the electromagnetic properties of Pb isotopes are carried out. The results are compared with available experimental data and are discussed in relation to the occupancy of the $0i_{13/2}$ orbital. We hope it can be a useful guide for the extensive experimental investigation underway and for our eventually clarification of the role played by nuclear deformation in this region. 

\section*{Acknowledgement}

This work was supported by the Swedish Research Council (VR) under grant Nos. 621-2012-3805, and 621-2013-4323 and from the National Natural Science Foundation of China No. 11405109. The calculations were performed on resources provided by the Swedish National Infrastructure for Computing (SNIC) at PDC at KTH, Stockholm.

\end{document}